# Techno-Economic Assessment in Communications: Models for Access Network Technologies

Carlos Bendicho
Independent ICT Researcher, ACM Member, IEEE Communications Society Member, FITCE Member, Spain
e-mail: carlos.bendicho@coit.es

*Abstract*—This article shows State of the Art of techno-economic modeling for access network technologies, presents the characteristics a universal techno-economic model should have, and shows a classification and analysis of techno-economic models in the literature based on such characteristics. As a result of his research in this direction, the author created and developed a Universal Techno-Economic Model and the corresponding methodology for techno-economic assessment in multiple domains, currently available for industry stakeholders under specific licence of use.

## I. Introduction

The development and evolution of Software Defined Networking (SDN) and Network Function Virtualization (NFV) technologies have created a myriad of different SD-WAN vendor solutions. Besides, some telecom operators are offering SD-WAN as a service. SD-WAN adoption rate is still slow but current context requires organizations to evolve WAN networks towards SD-WAN leveraging enhanced monitoring capabilities, QoS and Security policies application and Virtual Network Functions (VNF) execution either from Customer Premises Equipment (CPE) or from any other point in network as for example a datacenter.

All this increases technical complexity and makes it difficult for organizations to take decision about choosing a SD-WAN solution as they need an effective techno-economic assessment of different SD-WAN solutions.

Based on the author´s doctoral dissertation [82], this paper presents State of the Art of Techno-Economic Assessment in Communications, focusing on models for access network technologies.

Techno-economic models in the literature are based on the traditional definition of techno-economic model as "method for evaluating the economic feasibility of complex technical systems," according to the thesis of Smura [1].

Regarding the origin of the techno-economic modeling, Smura writes [1]:

*"The nature of techno-economic modeling and analysis is usually future oriented and uses and combines a number of methods from the field of Future-Oriented Technology Analysis (FTA). Among these methods is the cost-benefit analysis, scenario analysis, trends, expert panels and quantitative modeling (for an exhaustive list of other families and FTA methods, see TFAMWG, 2004, and [Scapolo & Porter, 2008, p . 152]). Although these methods and their combinations have been widely used by both academics and practitioners, academic work under the term "techno-economic" (eg .: modeling, analysis, evaluation, assessment) has mainly been published related to energy (eg [Zoulias & Lymberopoulos 2006]), biotechnology (eg .: [Hamelinck et al., 2005]) and telecommunications (ej.:[2]) especially by European research groups.*

*In the context of telecommunications, the term 'techno-economic' was introduced during the European research programme RACE (Research into Advanced Communicacions for Europe) in 1985-1995. Early techno-economic modeling work was done in the RACE 1014 ATMOSPHERIC project [7] [8] [9] and the RACE 1044 project [10] where alternative scenarios and strategies for evolution towards broadband systems were analysed. Later, the RACE 2087 TITAN (Tool for Introduction scenarios and Techno-economic studies for the Access Network) project developed a methodology and a tool for techno-economic evaluation of new narrowband and broadband services and access networks (see [2] [11]). Since the late 1990s, many European research projects have used and extended the methodologies and tools created in the early projects. "*

Note the following exception found in the literature to the traditional definition of a techno-economic model indicated by Smura in 2012. In 2010, [12] states: "Every business modeling should be accompanied by a technical and economic assessment so as to provide the reader with information on the financial perspective and the technical feasibility of a proposed investment project in telecommunications". The mentioned reference introduced the need for performance analysis of the access network, considering cost and reliability, but limited to relate both aspects in an indicator or specific figure of merit. Therefore, [12] suggests the assessment of technical feasibility, but eventually does not develop it.

Following a review and detailed analysis of the models in the literature, they are imbued with the traditional definition of techno-economic model indicated by Smura [1] and are eminently oriented to deployment of access technologies from the perspective of telecom operators, manufacturers and standards organizations. Only some models have capacity for evaluation of different access technologies, and a few of them include evaluation for combination of access technologies. Only one model has shown a slight hint of guidance to end users or agents other than those mentioned. Furthermore, all output parameters are economical, except for some very exceptional wink.

We proceed to show the review and analysis of the literature of techno-economic modeling of access technologies and the timing of projects related to public funding, given the interest in this research area shown by the institutions of the European Union (EU).

## II. Review and analysis of the literature

We have reviewed and analyzed the literature of techno-economic assessment models for access technologies, finding that it is based on the aforementioned traditional concept of techno-economic model enunciated by Smura [1], with the outstanding suggestion already indicated, and not developed [12].

We have selected a representative sample of the most relevant articles in literature which allow to provide insight on the State of the Art of techno-economic models for access technologies. Those papers are listed below in Table I.

TABLE I. PAPERS THAT DEVELOP OR USE MODELS OF TECHNO-ECONOMIC EVALUATION

| Author, year | Title | Techno-economic model | Access Technologies |
|---|---|---|---|
| Reed & Sirbu (1989) [13] | 'An Optimal Investment Strategy Model for Fiber to the Home' | dynamic programming | FTTH |
| Lu et al. (1990) [14] | 'System and Cost Analyzes of Broad-Band Fiber Loop Architectures' | Cost modeling | B-ISDN, four alternative architectures for fiber loop (loop fiber) (ADS, PPL HPPL, PON) |
| Graff et al. (1990) [7] | 'Techno-Economic Evaluation of the Transition to Broadband Networks' | STEM | Evolution of STM to ATM |
| Ims et al. (1996) [11] | 'Multiservice Access nework Upgrading in Europe: A Techno-Economic Analysis' | TITAN | xDSL, FTTx, HFC, FTTH (PON) |
| Olsen et al. (1996) [2] | 'Techno-Economic Evaluation of Narrowband and Broadband Access Network Alternatives and Evolution Scenario Assessment' | TITAN | ADSL, PON, CATV, ISDN, FTTx, HFC |
| Ims et al. (1997) [15] | 'Risk Analysis of Residential Broadband upgrade in a Changing Market and Competitive' | TITAN | xDSL, HFC, ATM PON |
| Stordahl et al. (1998) [16] | 'Risk Analysis of Residential Broadband upgrade based on Market Evolution and Competition' | OPTIMUM (based on TITAN) | FTTN, FTTB, HFC |
| Jankovic et al. (2000) [17] | 'A Techno-Economic Study of Broadband Access Network Implementation Models' | P614 | ISDN, xDSL, HFC, FTTx, WLL, Satellite |
| Katsianis et al. (2001) [18] | 'The Financial Perspective of the Mobile Networks in Europe' | TERA | GPRS, UMTS |
| Welling et al. (2003) [19] | 'Techno-Economic Evaluation of 3G & WLAN Business Case Under Varying Conditions Feasibility' | TONIC | UMTS, WLAN |
| Smura (2005) [20] | 'Competitive Potential of WiMAX in the Broadband Access Market: A Techno-Economic Analysis' | based on ECOSYS / TONIC | WiMAX |
| Monath et al. (2005) [21] | 'Muse Techno-economics for fixed access network evolution scenarios - DA3.2p' | MUSE | FTTx, ADSL, SHDSL, VDSL, xDSL over Optics |
| Sananes et al. (2005) [22] | 'Techno-Economic Comparison of Optical Access Networks' | e-Photon / One | FTTH |
| Lahteenoja et al. (2006) [23] | 'ECOSYS "techno-economics of integrated communication systems and services". Deliverable 16: "Report on techno-economic methology"' | ECOSYS | ISDN, B-ISDN (FITL), xDSL, HFC, FTTx, WLL, Satellite, WiMAX |
| Olsen et al. (2006) [24] | 'Technoeconomic Evaluation of the Major Telecommunication Investment Options for European Players' | ECOSYS / TONIC | HFC, ADSL, VDSL, LMDS, satellite, 3G, WLAN, FTTC, FTTH, |
| Pereira (2007) [5] | 'A Cost Model for Broadband Access Networks: FTTx versus WiMAX' | Proprietary (BATET) | FTTx, WiMAX |
| Chowdhury et al. (2008) [25] | 'Comparative Cost Study of Broadband Access Technologies' | Proprietary | xDSL, cable modem, FTTx, WiFi, WiFi + Hybrid FTTx, Hybrid FTTx + WiMAX (WOBAN) |
| Pereira & Ferreira (2009) [3] | 'Access Networks for Mobility: A Techno-Economic Model for Broadband Access Technologies' | Proprietary (BATET) | Static layer: FTTH (PON), xDSL, HFC, PLC; Nomadica layer (mobile users): WiMAX |
| Van der Merwe et al. (2009) [26] | 'A Model-based Techno-Economic Comparison of Optical Access Technologies' | Proprietary | FTTH Optical Networks: GPON, AON / Active Ethernet (AE), P2P |

| Odling et al. (2009) [27] | 'The Fourth Generation Broadband Concept' | ECOSYS | FTTdp (G.fast) |
|---|---|---|---|
| Ghazisaidi & Maier (2009) [28] | 'Fiber-Wireless (Fiwi) Networks: A Comparative Analysis of Techno-Economic EPON and WiMAX' | Proprietary | FTTH + WiMAX |
| Verbrugge et al. (2009) [29] | 'White Paper: Practical Steps in Techno-Economic Evaluation of Network Deployment Planning' | OASE | FTTH |
| Casier et al. (2010) [30] | '"Overview of Methods and Tools" Deliverable 5.1. OASE ' | OASE | FTTH |
| Zagar & Krizanovic (2010) [31] | 'Analyzes and Comparisons of Technologies for Rural Broadband Implementation' | Proprietary (Rural Broadband in Croatia) | ADSL, WiMAX |
| Vergara et al. (2010) [32] | 'COSTA: A Model to Analyze Next Generation Broadband Access Platform Competition' | COSTA (based on BREAD & TONIC & MUSE) | FTTH / GPON, FTTH / VDSL, FTTH / P2P, HFC / DOCSIS, WiMAX, LTE |
| Chatzi et al. (2010) [33] | 'Techno-economic Comparison of Current and Next Generation Optical Access Networks Long Reach' | BONE | FTTH duplicated for reliability and FTTH ring WDM / TDM PON fibers (SARDANA architecture) |
| Rokkas et al. (2010) [34] | 'Techno-Economic Evaluation of FTTC / VDSL and FTTH Roll-Out Scenarios: Discounted Cash Flows and Real Option Valuation' | ECOSYS | FTTC / VDSL and FTTH |
| Casier et al. (2011) [35] | 'Techno-Economic Study of Optical Networks' | OASE | FTTH |
| Feijóo et al. (2011) [6] | 'An Analysis of Next Generation Access Networks Deployment in Rural Areas' | Proprietary (model costs) | FTTH (GPON), FTTC / FTTB / VDSL, HFC DOCSIS 3.0, LTE (4G) |
| Martin et al. (2011) [36] | 'Which Could be the Hybrid Fiber Coax role of Next Generation Access Networks in?' | Proprietary (model costs) | FTTH (GPON), HFC DOCSIS 3.0 |
| Machuca et al. (2012) [37] | 'Cost-based assessment of NGOA Architectures and Its Impact in the business model' | OASE | Wavelength-routed WDM PON, Ultra Dense WDM, PON, AON with WDM |
| Van der Wee et al. (2012) [38] | 'A modular and hierarchically structured techno-economic model for FTTH Deployments' | OASE | FTTH (PON) FTTH (AON) |
| Walcyk & Gravey (2012) [39] | 'Techno-Economic Comparison of Next-Generation Access Networks for the French Market' | BONE | xDSL, FTTH (GPON), FTTH (LROA-SARDANA) |
| Pecur (2013) [4] | 'Techno-Economic Analysis of Long-Tailed Hybrid Fixed Wireless Access' | Proprietary | FIWI (Fixed-Wireless); Fixed: xDSL, FTTx, FSO; Wireless: WiFi, WiMAX, LTE (4G) |
| Bock et al. (2014) [40] | 'Techno-Economics and Performance of Convergent Radio and Fiber Architectures' | TITAN cost analysis | Active Remote Node PON FTTH combining + Radio Base Station (architecture Sodales) |
| Moreira & Zucchi (2014) [41] | 'Techno-economic evaluation of wireless access technologies for network environments campi' | TONIC & ECOSYS | WiFi, WiMAX, LTE |
| Ruffini et al. (2014) [42] | 'DISCUS: An End-to-End Solution for Ubiquitous Broadband Optical Access' | OASE | FTTP |
| Katsianis & Smura (2015) [43] | 'A model cost data for radio access networks' | Proprietary | LTE |
| Forzati et al. (2015) [44] | 'Next-Generation Optical Access Seamless Evolution: Results of the European Concluding FP7 Project OASE' | OASE | FTTH |
| Van der Wee et al. (2015) [45] | 'Techno-Economic Evaluation of Open Access on FTTH Networks' | OASE | FTTH |
| Shahid & Mas (2017) [80] | 'Dimensioning and Assessment of Protected Converged Optical Access Networks' | Proprietary | Converged Access Networks (FTTB/FTTH //LTE) |
| Oughton et al. (2019) [81] | 'An Open-Source Techno-Economic Assessment Framework for 5G Deployment' | Proprietary | 5G |

From the literature review, it is found that there is an American seed in the field of techno-economic modeling for access networks, in the late 80s and early 90s. Specifically, in 1989, predictions were published regarding the most appropriate moment to invest massively in FTTH (Fiber To The Home) access technology deployment using dynamic programming [13], identifying possible paths for investment from a pure copper access network to a FTTH network through hybrid networks, concluding that the optimum time to launch a massive deployment would not be before 2010, considering prediction of costs, income and interest rates. In the 90s and starting from [14] and [7], studies begin to focus, as well as Smura commented in his doctoral dissertation [1], on the detailed cost analysis, out of the components with a 'bottom-up' approach but ignoring the end-user perspective, and always oriented to the deployment of access networks, in order to compare the economic feasibility of different technical alternatives and identify parts of the access network that have a greater contribution to costs, considering different scenarios of evolution of the access network, as well as changing patterns of demand. It is noteworthy that in the US they were contemplating FTTH scenarios [14], starting from ISDN Narrowband (integrated services digital network or N-ISDN) to the Broadband ISDN (B-ISDN: Broadband ISDN), with focus in the detailed cost analysis using learning curves for predicting component costs [46].

In the 90s, it also germinates techno-economic modeling for access networks in Europe, with the first European projects with public funding from the EU (European Union), also focused on the assessment of the costs and oriented towards evaluating technical alternatives for deployment and network evolution. We find highlighted in this European germination the STEM [7] model as the precursor of a more complete model TITAN [11] [15], including a costs prediction model based on the so-called extended learning curve [2], which provides greater accuracy to successive predictive models as OPTIMUM [16] and sets the basis to more complete models as TONIC [19] [47], all based on evolutions from TITAN and always oriented to choose the most appropriate alternative for access network deployment by telecom operators, with the aim also to promote standards and recommendations.

The stage of development for more complete techno-economic models, initiated and inspired by TITAN, begins its consolidation with ECOSYS model [20][48], which enhances traditional techno-economic modeling based on the calculation of economic indicators such as NPV (Net Present Value) with the DCF analysis (Discounted Cash flows), ROA (Real options analysis) inspired in the so-called financial options or futures, in order to improve the accuracy of economic output parameters, and allows the techno-economic evaluation of wireline, wireless, and mixed or hybrid technologies in different scenarios and geographical areas to cover [23] [24] [49].

As a result of this aforementioned consolidation with the ECOSYS model, a dissemination effect occurs beyond the projects with public funding from the EU, which is detected by identifying new proprietary models as the model [50] for PLC technology (Power Line Communications), [51] for optical networks, [52] for 3G-LTE, the BATET model [53][5], and its subsequent evolution distinguishing between fixed and nomadic layers, the latter for mobile users [3], identifying also general input parameters in which there is a slight orientation to end user incorporating bandwidth transmission and reception requirements. Outside the public funding of the EU, the COSTA proprietary model [32] arises modeling costs of access network and based on MUSE, an extension from TONIC for the whole access and aggregation network [21], which runs parallel to ECOSYS project. More proprietary models appear like [26] oriented to optical access technologies comparison, and [25], [28] for hybrid networks that combine FTTx and WiFi or WiMAX.

The consolidation and dissemination continues. Multiple papers targeting specific scenarios that rely on the ECOSYS model are published as [27] oriented to FTTdp (Fiber To The Distribution Point) within the framework of the 4GBB initiative leading to the current G.fast standard [34].

The BONE project emerges, aimed at a future European optical network incorporating cost modeling for optical networks in the field of access/metro networks, seeking long range optical networks architectures to provide high reliability [33][54][55][39]. Under the BONE project, the article mentioned at the beginning of this section [12] is published, in which the assessment of the technical feasibility is suggested, after introducing the need for performance analysis of the access/metro network, considering the relationship between cost and reliability of the network. As stated before, [12] did not finally develop the evaluation of technical feasibility, but only limited to relate cost and reliability in a specific indicator or figure of merit, in order to evaluate various technical alternatives for long range optical networks with different mechanisms to increase reliability (e.g.: duplication of fibers, duplication of fibers and OLTs, duplication of fiber-OLT-ONU).

Fostered by public funding, emerges OASE project [29], which proposes a methodology based on the 'Plan-Do-Check-Act' (PDCA) by Shewhart/Deming, and adapts and reformulates it as 'Scope-Model-Evaluate-Refine', as well as a modular design of techno-economic modeling that integrates models and auxiliary methods around TONIC as 'framework tool' [30]. OASE extends view from ECOSYS, designing a modular framework for techno-economic modeling with the above methodology, enabling top-down and bottom-up approaches in techno-economic evaluation of optical access networks, becoming a model of relevance [56] [35] [57] [58] [59] [37] [38] [60] [61] [62] [63] [44] [64] [45]. The OASE model is used by other European projects in the field of optical networks, as DISCUS [42].

As a result of the mentioned spread effect, more proprietary models arise such as [65], [66], [67], [TS13] and [SIL13] for optical networks, [31], [68], [69], [70] [6], [71] and [72] for deployment of broadband in rural areas, [36] comparing FTTH and HFC DOCSIS 3.0, [4] for hybrid networks FiWi (Fixed-Wireless) which particularly distinguishes between investors and lenders related to financial agents. Reference [43] to deploy LTE networks in Finland, models power consumption of radio access data networks as a function of data traffic to deploy wireless access points [73]. Studies are also published based on the TONIC and ECOSYS models for specific scenarios (e.g.: wireless networks on campus) [41]. Projects with public funding from the EU emerge, relying on these consolidated tools, in order to assess new technologies, such as the IST-Sodales project developing an Active Remote Node FTTH combining PON + Radio Base Station (SODALES architecture), which evaluates it techno-economically using the TITAN tool [40]. More recently, [80] aims at converged accessed networks combining FTTH/FTTB and LTE, and [81] for 5G deployment. As mentioned, models in literature are based on the traditional definition of techno-economic model indicated by Smura [1], and are mainly oriented towards deployment of access network technologies from the perspective of operators, manufacturers and standardization bodies.

Given the limited features that are detected in the literature regarding the ability of multiaccess comparison, combination of technologies, orientation to end user or other agents in the telecommunications sector, the market dynamics and the context already mentioned, we conclude that it is interesting to deepen into the characteristics that a theoretical, universal and generalizable techno-economic model should have, which will be discussed in section IV.

III. CHRONOLOGY OF PUBLICLY FUNDED PROJECTS

At European level, public institutions of the European Union, have promoted and financed various projects in the past two decades, aimed at developing models for techno-economic evaluation of access technologies from the pioneers RACE 1014 ATMOSPHERIC, RACE 1028 REVOLVE, IBC 1044 RACE, RACE TITAN 2087, OPTIMUM AC226, AC364 TERA, IST-25172 TONIC, through EURESCOM, MUSE, BREAD, ECOSYS, OASE, etc. projects [7][8] [74][23][75][76][1].

Publicly funded projects mentioned above give rise to much of the literature, as can be seen in Table I, since many of the techno-economic models carry the name of the project that defines them. Other publicly funded projects also arise with different but related objectives, for example, related to backhaul or backbone networks that use and rely on techno-economic models developed by previous or parallel projects [77][42][40].

Techno-economic evaluation scenarios of above projects are closely related to the evolution of access technologies. In Fig. 1 the historical evolution of access technologies is shown, along with the timing of projects that develop or use techno-economic models, subject of literature in relevant publications and conferences.

All publicly funded projects identified are European. No other publicly funded projects have been found in other continents. The literature references from other continents, on the other hand minority, come from private companies and some universities, as can be seen in Table I [13][14][28][4] and [79] related to 'Digital India' initiative that is based on a European model [78][32].

According to the above, and in light of the number of research projects in this area, funded by institutions of the European Union, there is an economic interest and public funding, which justifies further deepening into the development of techno-economic evaluation models for access technologies.

It is observed that, despite having developed techno-economic models for access technologies by publicly funded projects, up to date it has not been identified a universal model for comparing any access technologies in any configuration or combination; which is oriented to any agent of the telecommunications sector; that makes it possible evaluation and comparison of technical feasibility and not only economical one, and that is flexible, extensible and integrable with other techno-economic models.

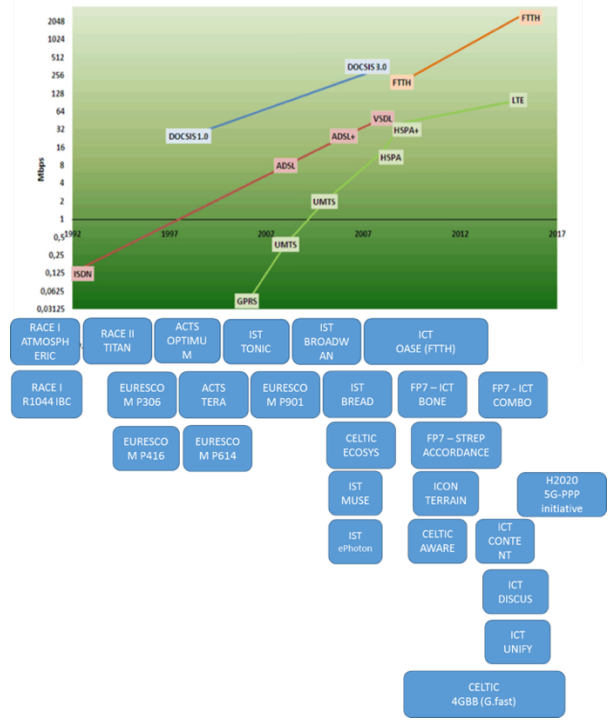

Figure 1. Historical evolution of projects that develop or use models of techno-economic assessment for access networks, along with the historical evolution of access technologies. Source: ICP-ANACOM, European Commission (ec.europa.eu) and websites of each project.

Therefore, the review and analysis of the literature, the existence of a wide variety of settings and access technologies, the high cost of investment and maintenance of the access network as well as the significant volume of scientific production supported by EU public funding for projects that promote and use techno-economic modeling, lead to the conclusion that it is interesting to deepen and lay the foundations regarding the characteristics required to have a universal and generalizable theoretical model for techno-economic evaluation of access technologies, in order to develop a specific classification and more accurately detect areas of improvement in techno-economic models of literature.

IV. CHARACTERISTICS OF A UNIVERSAL TECHNO-ECONOMIC MODEL

The features or characteristics a universal and generalizable theoretical model for techno-economic evaluation of access technologies should have are discussed below. The definition of these characteristics is based on a thorough study of the State of the Art, supplemented by author´s professional experience as a telecommunications engineer in designing innovative solutions in the field of access networks for different agents in the telecommunications sector:

- **Multiaccess Universality:** It must allow to compare different current and future access technologies.
- **Universality in Combination of technologies:** It must allow technical and economic assessment of accesses made by combining different technologies.
- **Universality in User orientation of technology:** It should be a model oriented to telecom operators and customers, telecommunications services end users, as well as to any other agent of the telecommunications market, such as regulators, Communication Service Providers (CSP) or technological consultancy firms.
- **Universality in incorporating "micro" and "macro" approaches:** It must incorporate "micro" ('bottom-up') approach (from the perspective of customer or end user) and "macro" approach ('top-down') (from the perspective of the deployment), when assessing techno-economically access technologies.
- **Orientation to User of the Model Requirements:** It must consider the requirements by the user of the model, be this customer, operator or any other agent to appraise access technologies. This feature is related to the model methodology of application.
- **Geographical Universality:** It must allow its application in any geographic area or geotype, whatever its population density, population segments (households, businesses) and distribution are.
- **Technical and Economic Universality:** It must provide technical and economic input parameters and technical and economic output, in order to allow both technical and economic assessment of different access technologies.
- **Extensibility and Flexibility:** the model must be extensible and flexible. It should provide easiness to add new input and output parameters contributing to its universality.
- **Technical and Economic Comparability:** It must make it easy to compare both technical and economic results with other models.
- **Predictive ability:** It must allow to incorporate and make predictions for a certain period of time.
- **Ability to integrate with other models:** It must allow integration with other techno-economic models to favor the most complete assessment possible and facilitate the evolution of the Art.

V. CLASSIFICATION AND ANALYSIS OF TECHNO-ECONOMIC MODELS FOR ACCESS NETWORKS

Then we proceed to classify a selection of techno-economic models for access networks identified in the literature, based on the above characteristics for a universal and generalizable techno-economic model.

We select a sample of 16 articles out of the 42 listed in Table I. The models for classification are selected by choosing 8 articles corresponding to models from public projects with EU funding, and 8 related to proprietary models (for which public funding has not been identified).

Every characteristic is composed by several items in order to identify whether the model under study is compliant with each item and evaluate its degree of compliance with the whole characteristic. We list items considered for each characteristic in Table II.

TABLE II. ITEMS CONSIDERED FOR EACH CHARACTERISTIC

| Characteristic | Items |
|---|---|
| Multiaccess Universality | -Fixed Access Technologies<br>-Wireless Access Technologies<br>-Mixed Access Technologies (Hybrid) |
| Universality in Combination of technologies | -Series Combination of Fixed Technologies<br>-Series Combination of Fixed and Wireless Technologies<br>-Parallel Combination of Different Technologies |
| Universality in User orientation of technology | -Oriented to Telcos (deployment KPIs)<br>-Oriented to Customers (Usage KPIs)<br>-Oriented to Other Agents |
| Universality in incorporating "micro" and "macro" approaches | -Incorporates "micro" approach (end user perspective)<br>-Incorporates "macro" approach (telco deployment perspective) |
| Orientation to User of the Model Requirements | -Economical Requirements<br>-Technical Requirements |
| Geographical Universality | -Geographical Area Description<br>-Existing infrastructure<br>-Population Mix Description |
| Technical and Economic Universality | -Input Technical Parameters<br>-Input Economical Parameters<br>-Output Technical Parameters<br>-Output Economical Parameters |
| Extensibility and Flexibility | -Easiness to add new input parameters<br>-Easiness to add new output parameters |
| Technical and Economic Comparability | -Technical Output Comparability<br>-Economical Output Comparability |
| Predictive Ability | -Period of Study as input parameter<br>-Allows input parameters with time prediction |

| | -Produces time prediction for output parameters |
| --- | --- |
| Ability to integrate with other models | -Allows to integrate as input other model output<br>-Model logic allows to incorporate easily other models parameters |

### A. Overall rating and ranking

In order to assess for each selected model the degree of compliance with the set of characteristics that is considered for a universal, scalable, flexible and generalizable techno-economic model, we use the following method, considering, for simplicity, that all items that compose a characteristic have the same weight:

- We assign '1' value for each item of a characteristic when the model under study is compliant with such item, and assign '0' value if the model is not compliant.
- We calculate the evaluation of each characteristic as the total sum of values for items that compose it.
- The total valuation for each model is the sum of the evaluations of all characteristics.

The result is shown in Table III. Normalizing for each feature in base 100, we obtain the degree of compliance of each model of the literature regarding the maximum possible score.

The maximum score in 2016, when the author elaborated this study for first time, corresponded to [3] with a fulfillment of 56%, thereby identifying a gap of 44 points to 100%, showing the opportunity to deepen and research in the development of proposals that reach a higher degree of compliance. After the last update of the literature review in 2020, the máximum score corresponds to [80] with a fulfillment of 62% which shows an improvement but still leaving a gap of 38 points.

The ranking of models presented in Table III shows, taking as reference the models [80] and [3] with greater compliance, that the path of improvement must focus on the following characteristics:

- Universality in Combination of access technologies
- Universality in User orientation
- Universality in incorporating "micro" and "macro" approaches
- Orientation to User of the Model Requirements ([80] reached compliance but not the rest)
- Technical and Economic Universality ([80] reached compliance but not the rest)
- Flexibility and Extensibility
- Technical and Economic Comparability ([80] reached compliance but not the rest)
- Ability to integrate with other models

### VI. CONCLUSIONS

In this paper, we have shown:

- in section II, a review and analysis of the literature
- in section III, a chronology of public projects with EU funding that develop and / or use models of techno-economic evaluation.
- in section IV, the characteristics of a theoretical universal, scalable, flexible and generalizable techno-economic model for access technologies are discussed.
- in section V, a classification of techno-economic models of the literature is made, based on the characteristics of the theoretical universal and generalizable techno-economic model set out in section IV. A ranking of techno-economic models of the literature is presented based on this classification.

After the review and classification of literature, it has been found that all models are oriented to deployment from the operator's perspective, and none is oriented to the end user, except for some exceptional wink in [3] which includes as input parameter the minimum transmission and reception bandwidth, and [5] that talks about QoS and a concurrency factor. All of them incorporate the "macro" approach from the perspective of deployment, but none incorporates the "micro" approach (end-user perspective).

Up to 2016, no model developed and provided technical output parameters, except BONE [12] suggesting an analysis of network performance, that eventually did not develop. However, in this last update after 4 years, two additional models have been identified with technical output parameters: [81] that includes capacity, coverage and energy efficiency as output parameters, and [80] which includes power consumption, availability and length of fiber related output parameters. Therefore, only these last two models of the literature allow the assessment of the technical performance of access technologies, while the rest lack technical comparability, in line with the traditional concept of techno-economic model stated by Smura [1].

Less than half of the models in the ranking, address a series combination of fixed access technologies. No model addresses the parallel combination of the same access technologies to improve technical performance. Eventually, in 2017, [80] addresses the parallel combination of different access technologies, in order to increase technical performance of the equivalent access, although there was slight glimmer in [2] with HFC (CATV) in parallel with TPON by 1996. Only [3] and [4] include the series combination of fixed + wireless technologies.

No model includes technical and economic requirements of the model user except [80], although oriented only to telecom operators. There is some slight hint of incorporating input information in [3] (minimum transmission and reception bandwidth), [5] (QoS and concurrency factor) and [6] which includes the Guaranteed Data Rate per User. However, no other model develops further this feature, incorporating, for example, a catalog or matrix of technical and economic requirements, as they all are oriented to the deployment of access technologies by telecom operators.

No model is identified that allows to add new input and output parameters in a flexible and simple way, so it is concluded that they are not flexible and extensible, probably motivated by the fact that all focus mainly on the assessment of economical viability.

No model includes the default logic for incorporation of other models parameters, thus limiting their ability to integrate with others. On the other hand, as stated before, all models but recent [80] and [81], aim to evaluate only the economical feasibility, lacking the assessment of technical feasibility.

Therefore, the review, classification and analysis of the literature shows that there is currently room for improvement in order to develop models that meet the characteristics of a flexible, generalizable and scalable universal techno-economic model, which allow analysis and comparison of network access technologies, and probably their extension and applicability to other domains.

As explained, it makes sense to deepen and research in the development of models for techno-economic assessment of network access technologies to achieve a higher degree of overall compliance, and in each of the characteristics, thus approaching the theoretical universal and generalizable techno-economic model.

Such universal and generalizable models could also be used for assessment in other domains beyond network access technologies as, for example, to satisfy the aforementioned needs of SD-WAN solutions techno-economic assessment.

As a result of his research in this direction, the author created and developed a Universal Techno-Economic Model (UTEM) and the corresponding methodology to use it for techno-economic analysis, assessment and decision-making in multiple domains. Table IV shows author´s model reaches an overall compliance of 92% as validated in [82].

This model is currently available for industry stakeholders under specific license of use.

TABLE III. OVERALL RATING AND RANKING OF LITERATURE DEPENDING ON THE DEGREE OF COMPLIANCE WITH THE CHARACTERISTICS OF A THEORETICAL UNIVERSAL, GENERALIZABLE, SCALABLE AND FLEXIBLE TECHNO-ECONOMIC MODEL (COMPLIANCE NORMALIZED BY FEATURE IN 100 BASIS)..

| | Multiaccess Universality | Universality in combination of access technologies | Universality in user orientation | Universality in incorporating "micro" and "macro" approaches | Orientation to User of the Model Requirements | Geographic universality | Technical and Economic Universality | Flexibility and Extensibility | Technical and Economic Comparability | Predictive Ability | Ability to integrate with other models | ASSESSMENT | %FULFILLMENT |
|---|---|---|---|---|---|---|---|---|---|---|---|---|---|
| Maximum possible score | 100 | 100 | 100 | 100 | 100 | 100 | 100 | 100 | 100 | 100 | 100 | 1100 | 100% |
| Shahid & Machuca [80] (2017) | 100 | 25 | 67 | 50 | 100 | 100 | 100 | 0 | 75 | 67 | 0 | 684 | 62% |
| Pereira & Ferreira [3] (2009) | 100 | 25 | 67 | 50 | 50 | 100 | 75 | 0 | 50 | 100 | 0 | 617 | 56% |
| Pereira [5] (2007) | 100 | 0 | 67 | 50 | 50 | 100 | 75 | 0 | 50 | 100 | 0 | 592 | 54% |
| Olsen et al. ECOSYS [24] (2006) | 67 | 25 | 34 | 50 | 0 | 100 | 75 | 0 | 50 | 100 | 50 | 551 | 50% |
| Monath et al. [21]. MUSE (2005) | 67 | 25 | 34 | 50 | 0 | 100 | 75 | 0 | 50 | 100 | 50 | 551 | 50% |
| Oughton et al. ]81] (2019) | 34 | 0 | 34 | 50 | 0 | 100 | 100 | 0 | 75 | 100 | 50 | 543 | 49% |
| Feijoo et al. [6]. RURAL (2011) | 67 | 25 | 34 | 50 | 50 | 100 | 50 | 0 | 50 | 100 | 0 | 526 | 48% |
| Vergara et al. [32]. Model COSTA (2010) | 67 | 25 | 67 | 50 | 0 | 100 | 50 | 0 | 50 | 100 | 0 | 509 | 46% |
| Olsen et al. [2]. TITAN (1996) | 34 | 50 | 34 | 50 | 0 | 100 | 50 | 0 | 50 | 100 | 0 | 468 | 43% |
| Jankovich et al. [17]. EURESCOM (2000) | 67 | 25 | 34 | 50 | 0 | 100 | 50 | 0 | 50 | 100 | 0 | 476 | 43% |
| Smura [20]. WiMAX only. TONIC & ECOSYS (2005) | 34 | 0 | 34 | 50 | 0 | 100 | 75 | 0 | 50 | 100 | 50 | 493 | 45% |
| Zagar et al. [31] (rural broadband in Croatia) (2010) | 67 | 0 | 34 | 50 | 0 | 100 | 75 | 0 | 50 | 100 | 0 | 476 | 43% |
| Pecur [4] FIWI (2013) | 100 | 25 | 0 | 50 | 0 | 100 | 75 | 0 | 50 | 67 | 0 | 467 | 42% |
| Martin et al. [36]. Only HFC (2011) | 34 | 0 | 34 | 50 | 0 | 100 | 50 | 0 | 50 | 100 | 0 | 418 | 38% |
| Van der Wee et al. [38]. FTTH only. OASE (2012) | 34 | 0 | 34 | 50 | 0 | 100 | 50 | 0 | 50 | 100 | 0 | 418 | 38% |
| Van der Merwe et al. [26]. FTTH only (2009) | 34 | 0 | 34 | 50 | 0 | 100 | 75 | 0 | 50 | 67 | 0 | 410 | 37% |

TABLE IV. OVERALL RATING OF AUTHOR´S MODEL AND RANKING OF LITERATURE DEPENDING ON THE DEGREE OF COMPLIANCE WITH THE CHARACTERISTICS OF A THEORETICAL UNIVERSAL, GENERALIZABLE, SCALABLE AND FLEXIBLE TECHNO-ECONOMIC MODEL (COMPLIANCE NORMALIZED BY FEATURE IN 100 BASIS)..

| | Multiaccess Universality | Universality in combination of access technologies | Universality in user orientation | Universality in incorporating "micro" and "macro" approaches | Orientation to User of the Model Requirements | Geographic universality | Technical and Economic Universality | Flexibility and Extensibility | Technical and Economic Comparability | Predictive Ability | Ability to integrate with other models | ASSESSMENT | %FULFILLMENT |
|---|---|---|---|---|---|---|---|---|---|---|---|---|---|
| **Maximum possible score** | 100 | 100 | 100 | 100 | 100 | 100 | 100 | 100 | 100 | 100 | 100 | 1100 | 100% |
| **Author´s UTEM model – Bendicho (2016)** | 100 | 100 | 84 | 100 | 100 | 100 | 100 | 50 | 75 | 100 | 100 | 1009 | 92% |
| **Shahid & Machuca [80] (2017)** | 100 | 25 | 67 | 50 | 100 | 100 | 100 | 0 | 75 | 67 | 0 | 684 | 62% |
| **Pereira & Ferreira [3] (2009)** | 100 | 25 | 67 | 50 | 50 | 100 | 75 | 0 | 50 | 100 | 0 | 617 | 56% |
| **Pereira [5] (2007)** | 100 | 0 | 67 | 50 | 50 | 100 | 75 | 0 | 50 | 100 | 0 | 592 | 54% |
| **Olsen et al. ECOSYS [24] (2006)** | 67 | 25 | 34 | 50 | 0 | 100 | 75 | 0 | 50 | 100 | 50 | 551 | 50% |
| **Monath et al. [21]. MUSE (2005)** | 67 | 25 | 34 | 50 | 0 | 100 | 75 | 0 | 50 | 100 | 50 | 551 | 50% |
| **Oughton et al. ]81] (2019)** | 34 | 0 | 34 | 50 | 0 | 100 | 100 | 0 | 75 | 100 | 50 | 543 | 49% |
| **Feijoo et al. [6]. RURAL (2011)** | 67 | 25 | 34 | 50 | 50 | 100 | 50 | 0 | 50 | 100 | 0 | 526 | 48% |
| **Vergara et al. [32]. Model COSTA (2010)** | 67 | 25 | 67 | 50 | 0 | 100 | 50 | 0 | 50 | 100 | 0 | 509 | 46% |
| **Olsen et al. [2]. TITAN (1996)** | 34 | 50 | 34 | 50 | 0 | 100 | 50 | 0 | 50 | 100 | 0 | 468 | 43% |
| **Jankovich et al. [17]. EURESCOM (2000)** | 67 | 25 | 34 | 50 | 0 | 100 | 50 | 0 | 50 | 100 | 0 | 476 | 43% |
| **Smura [20]. WiMAX only. TONIC & ECOSYS (2005)** | 34 | 0 | 34 | 50 | 0 | 100 | 75 | 0 | 50 | 100 | 50 | 493 | 45% |
| **Zagar et al. [31] (rural broadband in Croatia) (2010)** | 67 | 0 | 34 | 50 | 0 | 100 | 75 | 0 | 50 | 100 | 0 | 476 | 43% |
| **Pecur [4] FIWI (2013)** | 100 | 25 | 0 | 50 | 0 | 100 | 75 | 0 | 50 | 67 | 0 | 467 | 42% |
| **Martin et al. [36]. Only HFC (2011)** | 34 | 0 | 34 | 50 | 0 | 100 | 50 | 0 | 50 | 100 | 0 | 418 | 38% |
| **Van der Wee et al. [38]. FTTH only. OASE (2012)** | 34 | 0 | 34 | 50 | 0 | 100 | 50 | 0 | 50 | 100 | 0 | 418 | 38% |
| **Van der Merwe et al. [26]. FTTH only (2009)** | 34 | 0 | 34 | 50 | 0 | 100 | 75 | 0 | 50 | 67 | 0 | 410 | 37% |